\documentclass[prl,aps,showkeys,showpacs]{revtex4}
\newcommand{\bfm}[1]{\mbox{\boldmath${#1}$}}
\begin{document}
\title{Nonlinear Schr\"odinger Equations within\\ the Nelson Quantization Picture}
\author {G. Kaniadakis}\email{giorgio.kaniadakis@polito.it}
\author {A.M. Scarfone}\email{antonio.scarfone@polito.it}
\affiliation{Dipartimento di Fisica and Istituto Nazionale di
Fisica della Materia\\  Politecnico di Torino, Corso Duca degli
Abruzzi 24, 10129 Torino, Italy.}
\date{\today}
\begin {abstract}
We present a class of nonlinear Schr\"odinger equations (NLSEs)
describing, in the mean field approximation, systems of
interacting particles. This class of NLSEs is obtained
generalizing expediently the approach proposed in Ref. [G.K.
Phys. Rev. A \bfm{55}, 941 (1997)], where a classical system
obeying to an exclusion-inclusion principle is quantized using
the Nelson stochastic quantization. The new class of NLSEs is
obtained starting from the most general nonlinear classical
kinetics compatible with a constant diffusion coefficient
$D=\hbar/2\,m$. Finally, in the case of $s$-stationary states, we
propose a transformation which
linearizes the NLSEs here proposed.\\
\end {abstract}
\pacs{05.30.-d, 03.65.-w, 05.20.-y} \keywords{Nonlinear
Schr\"odinger equation, Stochastic quantization method.}
\maketitle
One among the most investigated topics in quantum mechanics
regards the nonlinear generalizations of the Schr\"odinger
equation. In the last decades many nonlinear extensions of the
Schr\"odinger equation have been proposed in literature either to
explore the fundamental aspects of quantum mechanics, with the
usual linear theory representing only a limiting case, or to
describe particular physical phenomenologies.\\ Among the many
nonlinear variants of Schr\"odinger equations proposed in
literature, we recall the Bialynicki-Birula and Mycielski
\cite{Bialynicki} in which appears the term $-b\,\ln|\psi|^2\,$,
the Guerra and Pusterla model \cite{Guerra} introducing the
nonlinearity $\Delta |\psi|/|\psi|$ in order to save the
superposition principle of the linear quantum mechanics, and
finally, the Weinberg theory \cite{Weinberg} which proposes the
introduction of homogeneous
nonlinear terms in order to save, only partially, the superposition principle.\\
Many of the NLSEs proposed contain complex nonlinearities. For
instance, the Doebner-Goldin equation \cite{Doebner1,Doebner2},
belonging to the Weinberg class, were introduced as the most
general NLSE compatible with the Fokker-Planck equation for the
probability density.

In addition, a large number of NLSEs have been introduced in
order to study some phenomenologies in condensed matter physics.
In Ref.s \cite{Gagnon, Gagnon1} a NLSE with a nonlinearity of the
type $a_1\,|\psi|^2\,\psi+a_2\,|\psi|^4\,\psi+
i\,a_3\,\partial_x(|\psi|^2\,\psi)+(a_4+i\,a_5)\,\partial_x|\psi|^2\,\psi$,
is introduced to describe a single mode wave propagation in a Kerr
dielectric guide. Another example is found in Ref.
\cite{Malomed1} where the nonlinearity
$a_1\,|\psi|^2\,\psi+i\,a_2\,\psi+
i\,a_3\,\partial_{xx}\psi+i\,a_4\,|\psi|^2\,\psi$ takes into
account pumping and dumping effects of the nonlinear media and
can be used to describe dynamical modes of plasma physics,
hydrodynamics and solitons in optical fibers (Ref.
\cite{Malomed3} and references therein). Finally, NLSEs are used
to describe propagation of high power optical pulses in
ultrashort soliton communication systems \cite{Doktorov},
incoherent solitons \cite{Bang1}, and others.

In Ref. \cite{Kaniadakis}, starting from a classical system
obeying to an exclusion-inclusion principle, a NLSE with complex
nonlinearity was derived by using a direct generalization of the
stochastic quantization method proposed by Nelson \cite{Nelson}.
The classical nonlinear kinetics imposed by the
exclusion-inclusion principle was studied in Ref.s \cite{KQ,KB}.
Successively in Ref. \cite{KLQ} the problem of the most general
nonlinear classical kinetics was investigated. This last one
contains as a particular case the kinetics found in different
physical systems, like among others, the Bosons and Fermions
(generalized exclusion-inclusion principle), the quons and the
system of particles obeying to the Haldane statistics
\cite{KLQ1}. The above nonlinear kinetics can be used to describe
several physical phenomenologies like the Bose-Einstein
condensation, the superfluidity and superconductivity and so on.

In the present paper we discuss the problem concerning the
quantization of a classical system obeying to the most general
nonlinear kinetics compatible with a constant diffusional
coefficient, by using the method of Ref. \cite{Kaniadakis}. We
recall that in Ref. \cite{Mann} it has been presented another
generalization of the stochastic quantization method which
permits to obtain NLSEs containing only real nonlinearities.

Let us consider a classical stochastic Marcoffian process in
$n-$dimensional space $\cal R$, described by the following
evolution equation:
\begin{equation}
\frac{\partial\,\rho(t,{\bfm x})}{\partial\,t}=\int\limits_{\cal
R}\left[\pi(t,{\bfm y}\rightarrow{\bfm x})-\pi(t,{\bfm
x}\rightarrow{\bfm y})\right]\,d^ny \ ,\label{ee}
\end{equation}
where $\rho(t,{\bfm x})$ is the particle density function and
$\pi(t,{\bfm x}\rightarrow{\bfm y})$ the transition probability
from the site $\bfm x$ to the site $\bfm y$ given by \cite{KQ}:
\begin{equation}
\pi(t,{\bfm x}\rightarrow{\bfm y})=r(t,{\bfm x},{\bfm x}-{\bfm y})
\, \alpha(\rho(t,{\bfm x}))\,\beta(\rho(t,{\bfm y})) \
.\label{tran}
\end{equation}
The above definition for  the transition probability implies a
nonlinear kinetics and plays a fundamental role in the
development of the theory exposed here. We observe that
$\pi(t,{\bfm x}\rightarrow{\bfm y})$ is given by a product of
three factors. The first one, $r(t,{\bfm x},{\bfm x}-{\bfm y})$,
is the transition rate and depends only on the starting  $\bfm x$
and arrival  $\bfm y$ sites during the particle transition ${\bfm
x}\rightarrow{\bfm y}$. The second factor $\alpha(\rho(t,{\bfm
x}))$ is an arbitrary function of the particle population of the
starting site. This function satisfies the condition
$\alpha(0)=0$, because if the starting site is empty the
transition probability is equal to zero. Finally the third factor
$\beta(\rho(t,{\bfm y}))$ is an arbitrary function of the arrival
site particle population. For this function we have the condition
$\beta(0)=1$ which requires that the transition probability does
not depend on the arrival site if, in it, particles are absent.
The expression of the function $\beta(\rho(t,\,\bfm{x}))$ plays a
very important role in the particle kinetics because can
stimulate or inhibit the transition ${\bfm x}\rightarrow{\bfm y}$
allowing in this way to take into account interactions originated
from collective effects.

The evolution equation (\ref{ee}), taking into account the
expression of the transition probability Eq. (\ref{tran}),
assumes the form:
\begin{eqnarray}
\nonumber \frac{\partial\,\rho(t,{\bfm
x})}{\partial\,t}&=&\beta(\rho(t,\,{\bfm x}))\,\int\limits_{\cal
R}r(t,\,{\bfm x}+{\bfm y},\,{\bfm y})\,\alpha(\rho(t,\,{\bfm
x}+{\bfm y}))\,d^ny\\
&-&\alpha(\rho(t,\,{\bfm x}))\,\int\limits_{\cal R} r(t,\,{\bfm
x},\,{\bfm y})\,\beta(\rho(t,\,{\bfm x}-{\bfm y}))\,d^ny \
.\label{ee1}
\end{eqnarray}
Now we consider the particle kinetics in the first neighbor
approximation and then we expand up to the second order the
quantities $r(t,\,{\bfm x}+{\bfm y},\,{\bfm
y})\,\alpha(\rho(t,\,{\bfm x}+{\bfm y}))$ and
$\beta(\rho(t,\,{\bfm x}-{\bfm y}))$ in Taylor series of ${\bfm
x}+{\bfm y}$  and ${\bfm x}-{\bfm y}$ respectively, in an
interval around $\bfm x$, for ${\bfm y}\ll{\bfm x}$. We obtain
\begin{eqnarray}
\nonumber r(t,\,{\bfm x}+{\bfm y},\,{\bfm
y})\,\alpha(\rho(t,\,{\bfm x}+{\bfm y}))&=&r(t,\,{\bfm x},\,{\bfm
y})\,\alpha(\rho(t,\,{\bfm x}))+\partial_{_i}[r(t,\,{\bfm
x},\,{\bfm y})\,\alpha(\rho(t,\,{\bfm
x}))]\,y_{_i}\\&+&{1\over2}\,\partial_{_i}\,\partial_{_j}[r(t,\,{\bfm
x},\,{\bfm y})\,\alpha(\rho(t,\,{\bfm x}))]\,y_{_i}\,y_{_j}+O(y^3)
\ ,\label{Taylor1}
\end{eqnarray}
\begin{eqnarray}
\beta(\rho(t,\,{\bfm x}-{\bfm y}))=\beta(\rho(t,\,{\bfm
x}))-\partial_{_i}\,\beta(\rho(t,\,{\bfm
x}))\,y_{_i}+{1\over2}\,\partial_{_i}\,\partial_{_j}\,\beta(\rho(t,\,{\bfm
x}))\,y_{_i}\,y_{_j}+O(y^3) \ ,\label{Taylor2}
\end{eqnarray}
being $\partial_{_i}=\partial/\partial\,x_{_i}$, with
$i=1,\,\cdots,\,n$.\\ After introducing the drift coefficient
${\bfm J}(t,{\bfm x})$ the diffusion coefficient $D(t,{\bfm x})$:
\begin{eqnarray}
{\bfm J}(t,{\bfm x})&=&\int\limits_{\cal R} {\bfm y}\,r(t,\,{\bfm
x},\,{\bfm y})\,d^ny \ ,\label{eej}\\ D(t,{\bfm
x})\,\delta_{_{ij}}&=&\frac{1}{2}\int\limits_{\cal R}
y_{_i}\,y_{_j}\,r(t,\,{\bfm x},\,{\bfm y})\,d^ny \ ,\label{ees}
\end{eqnarray}
Eq. (\ref{ee1}) transforms to the following nonlinear partial
differential equation \cite{KLQ,KLQ1}:
\begin{equation}
\frac{\partial\,\rho}{\partial\,t}={\bfm\nabla}\cdot\left[\left
({\bfm J}+{\bfm\nabla}\,D\right)\,\alpha\,\beta+D\,\left(\beta\,
\frac{\partial\,\alpha}{\partial\,\rho}-\alpha\,\frac{\partial\,
\beta}{\partial\,\rho}\right)\,{\bfm\nabla}\,\rho\right ] \
,\label{ce}
\end{equation}
with $\rho=\rho(t,\,{\bfm x}),\,\alpha=\alpha(\rho)$ and
$\beta=\beta(\rho)$. Eq. (\ref{ce}) is a continuity equation for
the density of particle $\partial\,\rho/\partial\,t
+{\bfm\nabla}\cdot{\bfm j}= 0$ in the configuration space, where
the current results to be the sum of two terms: ${\bfm j}={\bfm
j}_{\rm drift}+{\bfm j}_{\rm diff}$. The first one ${\bfm j}_{\rm
drift}=-({\bfm J}+{\bfm\nabla}\,D)\,\gamma(\rho)$ with
$\gamma(\rho)=\alpha(\rho)\,\beta(\rho)$, is a nonlinear drift
current. The second term ${\bfm j}_{\rm diff}=-D\,(\beta\,\partial
\alpha/\partial\,\rho-\alpha\,\partial\,\beta/\partial\,\rho
)\,{\bfm\nabla}\,\rho$ is a nonlinear diffusional current
different from the standard Fick current. In fact, the coefficient
which multiplies ${\bfm\nabla}\,\rho$ is a function of $\rho$.
Eq. (\ref{ce}) describes a class of nonlinear diffusional
processes varying the functions $\alpha(\rho)$ and $\beta(\rho)$.

In the following, we study a sub-class of the above processes for
which the diffusion coefficient $D$ is a constant and the
diffusional current is given by the standard Fick form ${\bfm
j}_{\rm diff}=-D\,{\bfm\nabla}\,\rho$. Then the functional
$\alpha$ and $\beta$ obey the condition:
\begin{eqnarray}
\beta\,\frac{\partial\,\alpha}{\partial\,\rho}-\alpha\,\frac{\partial\,
\beta}{\partial\,\rho}=1 \ .\label{def}
\end{eqnarray}
From Eq. (\ref{def}) we obtain:
\begin{equation}
\alpha(\rho)=\left[\gamma(\rho)\,\Gamma(\rho)\right]^{1/2} \
,\hspace{6mm}\beta(\rho)=\left[\gamma(\rho)/\Gamma(\rho)\right]^{1/2}\
\ ,\label{ab}
\end{equation}
where
\begin{equation}
\Gamma(\rho)=\exp\left[\int\limits^{\rho}{\frac{d\,\rho^\prime}
{\gamma(\rho^\prime)}}\right] \ .\label{g}
\end{equation}
Varying $\gamma(\rho)$, Eq.s (\ref{ce}), (\ref{ab}) and (\ref{g})
describe the class of nonlinear kinetics which is the object of
study of the present contribution. This class of kinetics includes
also the standard linear one for which $\gamma(\rho)=\rho$ and
then $\alpha(\rho)=\rho$, $\beta(\rho)=1$.

Following Nelson, we assume that the continuity equation
(\ref{ce}), which is a Fokker-Planck equation, describes a
stochastic Markoffian process asymmetric in time \cite{Nelson}.
For this reason we must consider both the forward $(+)$ and the
backward $(-)$ expressions of the particle current ${\bfm j}=
{\bfm u}^{(\pm)}\,\gamma(\rho)\mp D\,{\bfm\nabla}\,\rho$ where
${\bfm u}^{(\pm)}=-{\bfm J}^{(\pm)}$ are the forward and backward
velocities. The forward and backward Fokker-Planck equation
assume the form:
\begin{eqnarray}
\frac{\partial\,\rho}{\partial\,t}+{\bfm\nabla}\cdot\left[{\bfm
u}^{(\pm)}\,\gamma(\rho)\mp D\,{\bfm\nabla}\,\rho\right]=0 \ .
\end{eqnarray}
We define now the current velocity as ${\bfm v}=\left({\bfm
u}^{(+)}+{\bfm u}^{(-)}\right )/2$, assuming that can be derived
by a potential ${\bfm v}={\bfm\nabla}\,S/m$. After summing the
forward and backward Fokker-Planck equations, we obtain the
following continuity equation:
\begin{equation}
\frac{\partial\,\rho}{\partial\,t}+{\bfm\nabla}\cdot\left
[\frac{{\bfm\nabla}\,S}{m}\,\gamma(\rho)\right ]=0 \ .\label{ce1}
\end{equation}
In the stochastic quantization method the system is viewed as a
classical Markoffian process where the particles are subjected to
a Brownian diffusion with a diffusion coefficient given by
$D=\hbar/2\,m$ \cite{Fenyes,Weizel}. The mean acceleration for
this classical stochastic process is defined as
\begin{equation}
{\bfm a}=\frac{1}{2}\left[\left(\frac{d}{dt}\right)^{(-)}{\bfm
u}^{(+)}+\left(\frac{d}{dt}\right)^{(+)}{\bfm u}^{(-)}
 \right] \ ,\label{acc}
\end{equation}
where the forward and backward derivative are defined us:
\begin{equation}
\left(\frac{d}{dt}\right)^{(\pm)}=\frac{\partial}{\partial\,t}+
{\bfm u}^{(\pm)}\cdot{\bfm\nabla}\pm D\,\Delta \ .
\end{equation}
The dynamics is governed by the Newton law ${\bfm F}=m\,{\bfm a}$
and supposing that the mean external force ${\bfm F}$ is derived
from a potential ${\bfm F}=-{\bfm\nabla}\, V$, we can write
${\bfm\nabla}\,V=-m\,{\bfm a}$. This last equation, if we take
into account the expression of ${\bfm a}$ given by Eq. (\ref{acc})
and the expressions of the forward and backward velocities ${\bfm
u}^{(\pm)}={\bfm\nabla}\,S/m\pm D\,{\bfm\nabla}\,\rho/
\gamma(\rho)$, reproduces after integration the evolution
equation for the potential $S$ which is the following generalized
{\sl Hamilton-Jacobi} equation:
\begin{equation}
\frac{\partial\,S}{\partial\,t}+\frac{({\bfm\nabla}\,S)^2}{2\,m}
+U_q(\rho)+W(\rho)+V=0 \ ,\label{hj}
\end{equation}
where $U_q(\rho)=-\left(\hbar^2/2\,m\right)\,\Delta\,\rho^{1/2}/
\rho^{1/2}$ is the quantum potential \cite{Madelung,Bohm} and $W$,
given by
\begin{equation}
W(\rho)=f_1(\rho)\,\Delta\,\rho+f_2(\rho)\,({\bfm\nabla}\,\rho)^2
\ ,
\end{equation}
with
\begin{eqnarray}
&&f_1(\rho)=\frac{\hbar^2}{4\,m}\left[\frac{1}{\rho}
-\frac{1}{\gamma(\rho)}\right] \ , \\ &&f_2(\rho)=
\frac{\hbar^2}{4\,m}\left[\frac{1}{\gamma(\rho)^2}
\frac{\partial\,\gamma(\rho)}{\partial\,\rho}
-\frac{1}{2\,\gamma(\rho)^2}-\frac{1}{2\,\rho^2}\right] \ ,
\end{eqnarray}
is a real nonlinear quantity depending only on the density of
particle $\rho$.\\ We introduce now the complex function $\psi$
given by the ansatz:
\begin{eqnarray}
\psi(t,\,{\bfm x})=\rho(t,\,{\bfm x})
^{1/2}\exp\left[\frac{i}{\hbar}\,S(t,\,{\bfm x})\right] \
.\label{ans}
\end{eqnarray}
By taking into account Eq. (\ref{ans}) the particle current ${\bfm
j}=({\bfm\nabla}\,S/m)\,\gamma(\rho)$ can be written as:
\begin{equation}
{\bfm j}=-\frac{i\,\hbar}{2\,m}\,\gamma(\rho)\,\left
(\frac{{\bfm\nabla}\,\psi}{\psi}-\frac{{\bfm\nabla}\,\psi^\ast}{\psi^\ast}\right)
\ .\label{j}
\end{equation}
The continuity equation (\ref{ce1}), the generalized {\sl
Hamilton-Jacobi} equation (\ref{hj}) together with the ansatz
(\ref{ans}), allow to obtain the evolution equations for the
function $\psi$ which is the following NLSE:
\begin{equation}
i\,\hbar\,\frac{\partial\,\psi}{\partial\,t}=-\frac{\hbar^2}{2\,m}
\,\Delta\,\psi+V\,\psi+\Lambda(\rho,\,{\bfm j})\,\psi \
,\label{se}
\end{equation}
where $\Lambda(\rho,\,{\bfm j})= W(\rho)+i\,{\cal W} (\rho,\,{\bfm
j})$ is a nonlinear complex quantity whose imaginary part is
given by:
\begin{equation}
{\cal W}(\rho,\,{\bfm
j})=\frac{\hbar}{2\,\rho}\,{\bfm\nabla}\cdot\left
[\frac{\rho-\gamma(\rho)}{\gamma(\rho)}\,{\bfm j}\right] \
,\label{im}
\end{equation}
and the real one takes the form:
\begin{eqnarray}
W(\rho)=
\frac{\hbar^2}{4\,m}\,\left[\left({1\over\rho}-{1\over\gamma}\right)\,\Delta\,\rho
+\left({1\over\gamma^2}\,
\frac{\partial\,\gamma}{\partial\,\rho}-{1\over2\,\gamma^2}-{1\over2\,\rho^2}\right)\,
\left({\bfm\nabla}\,\rho\right)^2\right] \ .\label{eqq}
\end{eqnarray}
We remark that Eq. (\ref{se}) defines a large class of NLSEs
by varying the functional $\gamma(\rho)$.\\
In Ref. \cite{Kaniadakis}, with the purpose of introducing a
generalized inclusion-exclusion principle which originates from
the collective effects of a many body quantum system, it has been
assumed $\gamma(\rho)=\rho\,(1+\kappa\rho)$. The corresponding
NLSE obtained from Eq. (\ref{se}) becomes:
\begin{eqnarray}
\nonumber
i\,\hbar\,\frac{\partial\,\psi}{\partial\,t}&=&-\frac{\hbar^2}{2\,m}\,
\Delta\,\psi+V\,\psi-i\,\kappa\,\frac{\hbar}{2\,\rho}\,{\bfm\nabla}
\cdot\left(\frac{\rho\,{\bfm
j}}{1+\kappa\,\rho}\right)\,\psi\\
&&+\kappa\,\frac{\hbar^2}{4\,m}\left[\frac{\Delta\,\rho}
{1+\kappa\,\rho}+\frac{2-\kappa\,\rho}{2\,\rho}\,
\frac{({\bfm\nabla}\,\rho)^2}{(1+\kappa\,\rho)^2}\right]\,\psi \
,\label{seip}
\end{eqnarray}
where $\bfm j$, given by Eq. (\ref{j}), takes the form:
\begin{eqnarray}
{\bfm j}=-\frac{i\,\hbar}{2\,m}\,(\psi^\ast\,{\bfm\nabla}\psi
-\psi\,{\bfm\nabla}\psi^\ast)\,(1+\kappa\,\rho) \ .\label{jeip}
\end{eqnarray}
In Eq.s (\ref{seip}) and (\ref{jeip}) the parameter
$\kappa\in{\bfm R}$ takes into account of exclusive ($\kappa<0$)
or inclusive ($\kappa>0$) effects due to the particle
interactions of the system.

Finally we consider the $s$-stationary states $({\bfm j}={\bfm
0})$ of the system described by Eq. (\ref{se}) for which ${\cal
W}(\rho,\,{\bfm 0})=0$. The field $\psi$ for these states has the
form $\psi=\rho^{1/2}\exp (-i\,E\,t/\hbar)$ and obeys the
stationary NLSE:
\begin{equation}
 E\,\psi =-\frac{\hbar
^2}{2\,m}\,\Delta\,\psi+V\,\psi+W(\rho)\,\psi \ .\label{nlse}
\end{equation}
Eq. (\ref{nlse}) describes a $C$-integrable system which is
linearizable by means of a transformation on the field $\psi$
\cite{Calogero,Calogero2}. In fact, if we introduce now the new
stationary wave function $\phi$ by means of the ansatz
\begin{equation}
\phi=c\,\Gamma(\rho)^{1/2}\,\exp (-i\,E\,t/\hbar) \ ,\label{an}
\end{equation}
from Eq. (\ref{nlse}) follows that the new field $\phi$ obey to
the stationary linear Schr\"odinger equation
\begin{equation}
E\,\phi=-\frac{\hbar^2}{2\,m}\,\Delta\,\phi+V\,\phi \ .\label{lse}
\end{equation}
The constant $c$ in Eq. (\ref{an}) is obtainable from the
condition $\int_{_{\cal R}} \psi^\ast\,\psi\,d^nx= \int_{_{\cal
R}}\phi^\ast\,\phi\,d^nx=1$ so that the wave-functions $\psi$ and
$\phi$ can be normalized simultaneously. The study of the
$s$-stationary states of the nonlinear system obeying to Eq.
(\ref{nlse}) is reduced to the study of the $s$-stationary states
of the linear system (\ref{lse}) under the same external potential
$V({\bfm x})$ and with the same energy $E$. We remark that the
linearization of Eq. (\ref{nlse}) can be understood by taking
into account that the wave-functions $\psi$ and $\phi$ satisfies
the relation  $(T\phi)/\phi=(T\psi)/\psi+W(|\psi|^2)$ where
$T=-(\hbar^2/2\,m)\,\Delta$ is the quantum operator of kinetic
energy. This last relationship means that the sum of the kinetic
energy and of the nonlinear term $W(|\psi|^2)$ is equal to the
kinetic energy of the system described by the wave function
$\phi$. This last system is linear because the nonlinear quantity
$W(|\psi|^2)$ is hidden into this kinetic energy.

In conclusion, in the present paper we have studied within
stochastic quantization a system obeying to the most general
nonlinear kinetics compatible with a constant diffusion. The
evolution equation (\ref{se}) obtained in this way describes a
wide class of NLSEs with complex nonlinearities by varying the
functional $\gamma(\rho)$. Being $\gamma(\rho)$ an algebraic
function of $\rho$ the nonlinearities are derivative of the second
order, with the real part depending on the density of particle
$\rho$ but not on the phases $S$. All equations belonging to this
class conserve the total number of particle of the system, they
are $U(1)$-invariant and, in absence of the external potential
$V({\bfm x})$, they are invariant over space-time translation and
rotation, but generally, are not Galilei invariant. Finally, in
the case of $s$-stationary states, a transformation
$\gamma$-dependent permits the linearization of this class of the
NLSEs.


\end{document}